\documentclass[11pt]{article}
\usepackage{amsmath,amssymb}
\usepackage{graphicx}

\begin{document}

\setcounter{page}{1}

\pagestyle{plain}

\begin{center}
\Large{\bf Observational Viability of Anisotropic Inflation Revisited}\\
\small \vspace{1cm} {\bf Maryam Roushan
$^{}$\footnote{m.roushan@umz.ac.ir }} \,,\quad {\bf Narges Rashidi
$^{}$\footnote{n.rashidi@umz.ac.ir (Corresponding Author) }}\,,
\quad {\bf Kourosh Nozari
$^{}$\footnote{knozari@umz.ac.ir }} \\
\vspace{0.5cm} Department of Theoretical Physics, Faculty of
Science,
University of Mazandaran,\\
P. O. Box 47416-95447, Babolsar, IRAN\\
\end{center}

\begin{abstract}
We investigate anisotropic inflation within the single-field model
featuring an intermediate scale factor. Our analysis reveals that
the anisotropic nature of the Friedmann equations in this framework
affects the slow-roll parameters, which in turn influence key
perturbation parameters. Using a numerical approach, we derive
constraints on the intermediate parameter $\beta$ and the
anisotropic parameter $c$. Our results show that the model is
consistent with Planck2018 TT, TE, EE +lowE+lensing+BK14+BAO data
at $68\%$ CL, for $0.84<\beta<1$ and $7.34<c<27.7$. At $95\%$ CL the
consistency holds for $0.77<\beta<1$ and $7.17<c<28.9$. The model is
also consistent with Planck2018 TT, TE, EE +lowE+lensing+BK18+BAO
data, for $0.91<\beta<1$ and $8.00<c<27.4$ (at $68\%$ CL), and
$0.88<\beta<1$ and $7.40<c<28.8$ (at $95\%$ CL). Additionally, we
examine the reheating phase using these constraints on constraints
on $\beta$ and $c$ and determine the observationally consistent
ranges for the number of e-folds and the temperature during the
reheating phase.
\end{abstract}
\newpage

\section{\label{sec1}Introduction}

To address some important shortcomings of the standard model of
cosmology, the intriguing inflationary paradigm has been
proposed~\cite{Gut81}. This framework offers an elegant explanation
for the dynamics of the universe during its earliest moments. In
Ref.~\cite{Lin82}, the author presents a new inflationary model that
resolves the issues of the old inflation model. In
Ref.~\cite{Alb82}, it was proposed that a more detailed examination
of first-order phase transitions in GUT models could help resolve
some of the biggest puzzles in cosmology. Observational tests of
cosmic inflation and the possibility of reconstructing the inflaton
potential based on data were examined in Ref.~\cite{Lid97}. More
significantly, inflation provides an explanation for how the
large-scale structure of the universe was initially seeded. The
simplest inflationary models rely on the slow-roll conditions, where
the inflaton field gradually rolls toward the minimum of its
potential. This straightforward single-field inflation model assumes
the cosmological principle, implying a homogeneous and isotropic
background geometry.

On the other hand, when we physically describe the very early
moments in the history of the universe, it may not be enough to just
consider a simple isotropic geometry. Also, from recent
observational data, it is revealed that there are some anomalies in
the cosmic microwave background radiation~\cite{Pl18b}. Some
cosmologists believe that these anomalies have instrumental
origins~\cite{Han09}. Also, the authors of ~\cite{Han10} suggest that some observed anomalies in the Cosmic Microwave Background (CMB) may result from systematic effects such as asymmetric beams. While, some other
cosmologists relate these anomalies to a homogeneous but anisotropic
universe~\cite{Ell69}. In this regard, Ref.~\cite{Gum07} investigates how inflationary perturbations in anisotropic backgrounds affect the CMB and how data can be used to identify these effects. Ref.~\cite{Pit08} also investigates the predictions from an anisotropic inflationary era and its effects on various cosmological features. There are various confrontations
to the anisotropy. According to the prediction of the cosmic no-hair
conjecture, the universe eventually approaches a homogeneous and
isotropic state, even if it started from an anisotropic
state~\cite{Gib77}. In Ref.~\cite{Sta83},
the local validity of this conjecture has been proved. In paper ~\cite{Mul90} it has been shown that the power-law inflation acts as a local attractor for inhomogeneous solutions in a model with a minimally coupled scalar field and a positive exponential potential. In Ref.~\cite{Bar84}  the role of initial conditions in inflationary cosmology and challenges the cosmic "no-hair" conjecture have been examined. Exact solutions of Einstein?s equations in a case with an inhomogeneous scalar field, resulting in an intrinsically inhomogeneous metric have been studied in Ref.~\cite{Ste87}. However, the
authors of Ref.~\cite{Col19} have explored the possibility of
anisotropy in the late-time universe. In Ref.~\cite{Al23}, one can
find a comprehensive review of FLRW anomalies. Some counterexamples
questioning the validity of the no-hair conjecture can also be found
in Ref.~\cite{Bar06a}. Also, the authors of~\cite{Bar06b} have explored Bianchi I and II universes in quadratic gravity, finding early-time isotropization and new anisotropic inflationary behaviors. Paper ~\cite{Mid10} presents new anisotropic solutions in higher-order gravity theories, exploring Bianchi type VIII, IX, and I cosmologies near the initial singularity and their relevance to cosmic ``no-hair" theorems. The authors of~\cite{Mul18} have studied the past evolution of an anisotropic Bianchi I universe in $f(R)$ gravity. Of course, according to
the studies in Ref.~\cite{Kao11a}, it seems that some
of these counterexamples are unstable during the inflationary phase
of the universe. Also, in Ref.~\cite{Kao09} it has been found that anisotropic solutions in a conformal-violating Maxwell model are unstable during inflation but can persist in a slowly expanding phase for certain parameters. The authors of Ref.~\cite{Mal12} proved an
inflationary cosmic no-hair theorem, which allows for a certain
level of anisotropy. In the supergravity-based model also, a
counterexample to the conjecture has been found which seems
interesting~\cite{Wat09}.

Therefore, by considering the fact that anisotropy may have an
important role in the history of the universe, some cosmologists
have performed interesting works on this issue. By incorporating
additional quadratic Ricci curvature terms into the Einstein-Hilbert
action, Ref.~\cite{Bar10} explores several cosmological solutions.
The power spectrum in anisotropic inflation has been examined in
Ref.~\cite{Dul10}, where a specific coupling between the gauge field
and the inflaton was considered by the authors. Ref.~\cite{Kan10}
presents exact solutions for anisotropic power-law inflation,
achieved by introducing exponential potentials for the inflaton and
the gauge field. Another interesting case has been studied in
Ref.~\cite{Lah16}. The study demonstrates anisotropic power-law
inflation, relying on exponential forms for the inflaton potential.
One can find the cosmological aspects of anisotropic
constant-roll inflation in Ref.~\cite{Ito18}. The properties of
scalar and tensor perturbations in the non-canonical anisotropic
inflation have been studied in Ref.~\cite{Do21a}. The authors of
Ref.~\cite{Do21b} derived an anisotropic power-law solution by
introducing two scalar fields which are non-minimally coupled to two
vector fields. A hyperbolic inflation model along a gauge field has
been considered in Ref.~\cite{Che21}, where the authors have
explored anisotropic inflation. Based on the Bianchi IX cosmology,
the authors of Ref.~\cite{Noj22} have considered the general form of
the metric and examined the anisotropic inflation in the
cosmological models with $F(R)$ gravity. Thus, it seems interesting
to consider anisotropic property in the cosmological model and study
its cosmological implication. In fact, the presence of the
anisotropic geometry may affect the cosmological viability of the
models. Particularly when the slow-roll parameters are influenced by
anisotropic effects, it would affect for instance the scalar
spectral index and the tensor-to-scalar ratio. By comparing the
numerical values of these parameters with observational data it is
possible to find some constraints on the anisotropic parameter.
Also, we can explore the presence of anisotropic geometry in the
reheating phase after inflation to find some extra information.

Now, in this paper, we consider single-field model with intermediate
scale factor in the anisotropic background and revisit the inflation
in this setup. Usually, anisotropic aspects seem
less relevant in the canonical scalar field model for inflation.
This is because the exponential expansion quickly dilutes any
initial anisotropy. However, here we focus on an intermediate
inflation model. Unlike standard exponential inflation, the
intermediate model features a slower expansion rate, which might
allow some anisotropy to persist. Our goal is to how it impacts the
model's viability. By using the slow-roll, perturbations
and reheating parameters, we try to check this anisotropic model
observationally. The structure of the paper is as follows. In
section \ref{sec2}, we study the anisotropic inflation and find some
important equations and parameters in this setup. In section
\ref{sec3}, we explore the observational viability of the
intermediate anisotropic inflation in comparison with the
observational data. The reheating phase after inflation is
considered in \ref{sec4}. In section \ref{sec5}, we present a
summary of the model.

\section{\label{sec2}Anisotropic Inflation}
To revisit the anisotropic inflation, we consider a homogeneous but
anisotropic background geometry defined by the following
metric~\cite{Noj22}
\begin{eqnarray}
\label{eq1}
ds^2=-dt^2+a(t)^{2}\,\sum_{i=1}^{3}\,e^{2\xi_{i}(t)}\,(dx^{i})^{2}\,.
\end{eqnarray}
In this metric, we defined
\begin{equation}
\bar{\xi}(t)=\frac{1}{3}\Sigma_{i=1}^{3}\xi_{i}(t),
\end{equation}
which gives the the average value of the parameter $\xi_{i}(t)$.
Following Ref.~\cite{Noj22}, it is possible to express
\begin{equation}a(t)\rightarrow a(t)+\bar{\xi}(t)
\end{equation}
 and
\begin{equation}
\xi_{i}(t)\rightarrow \xi_{i}(t)-\bar{\xi}(t).
\end{equation}
In this way, we find
\begin{equation}
\Sigma_{i=1}^{3}\xi^{i}=0
\end{equation}
 and
\begin{equation}
\Sigma_{i=1}^{3}\dot{\xi}^{i}=0.
\end{equation}
Therefore, we can find the components of the Ricci tensor
as~\cite{Ras2024}
\begin{eqnarray}
\label{eq2}
{\cal{R}}_{00}=-\sum_{i=1}^{3}(\dot{\xi}^{i})^2-3\dot{H}-3H^{2}\,,
\end{eqnarray}
\begin{eqnarray}
\label{eq3}
{\cal{R}}_{ij}=\left(\dot{H}+3H^2+\ddot{\xi}^{i}+3H\dot{\xi}^{i}\right)a^2e^{2\xi_{i}}\delta_{ij}\,.
\end{eqnarray}
Note that, in the equations of this paper, we use a dot to
demonstrate a cosmic time. We also find the Ricci scalar in the
anisotropic background as~\cite{Ras2024}
\begin{eqnarray}
\label{eq4}
{\cal{R}}=\sum_{i=1}^{3}(\dot{\xi}^{i})^2+6\dot{H}+12H^{2}\,.
\end{eqnarray}
In studying the inflation phase in every model, the Friedmann
equations play an important and critical role. To obtain these
equations in the anisotropic geometry situation, we use $(0,0)$ and
$(i,i)$ components of the Einstein field equations and find
\begin{eqnarray}
\label{eq5}
H^2=\frac{\kappa^{2}}{3}\left(\frac{1}{2}\,\dot{\phi}^{2}+
V(\phi)\right)+\frac{1}{6}\sum_{i=1}^{3}(\dot{\xi}^{i})^{2}\,,
\end{eqnarray}
\begin{eqnarray}
\label{eq6}
2\dot{H}+3H^{2}=-\kappa^{2}\,\left(\frac{1}{2}\,\dot{\phi}^{2}-
V(\phi)\right)-\frac{1}{2}\sum_{i=1}^{3}(\dot{\xi}^{i})^2\,,
\end{eqnarray}
where, $\kappa^{2}=8\pi G$ is the gravitational constant. Also, to
obtain these equation, we have used equations
(\ref{eq2})-(\ref{eq4}). By checking the equation of motion, we see
that the anisotropic doesn't have any effect on it and we reach the
usual formulae as
\begin{eqnarray}
\label{eq8}\ddot{\phi} +3H\dot{\phi}+V'=0\,.
\end{eqnarray}
From now on, we use a prime to show $\frac{d}{d\phi}$.

By considering the Friedmann equations, the potential in this model
is obtained as
\begin{eqnarray}
\label{eq9}V=\frac
{6\,{H}^{2}-\kappa^{2}\dot{\phi}^{2}-\sum_{i=1}^{3}(\dot{\xi}^{i})^2}{2\,{\kappa}^{2}}
\,,
\end{eqnarray}

with
\begin{eqnarray}
\label{eq9-2}
\dot{\phi}=\frac{\sqrt{-2\dot{H}-\sum_{i=1}^{3}(\dot{\xi}^{i})^2}}{\kappa}
\,.
\end{eqnarray}

At this point, from the definition of the slow-roll parameters
as $\epsilon=-\frac{\dot{H}}{H^{2}}$ and
$\eta=\frac{1}{H}\frac{\dot{\epsilon}}{\epsilon}$, and by using
equations (\ref{eq5})-(\ref{eq9}), we find these parameters in the
anisotropic inflation model as
\begin{eqnarray}
\label{eq10}\epsilon=\frac{1}{2\kappa^2}\frac{V'^{2}}{V^{2}}+\frac{3\sum_{i=1}^{3}(\dot{\xi}^{i})^2}{\kappa^{2}V+\frac{1}{2}\sum_{i=1}^{3}(\dot{\xi}^{i})^2}\,,
\end{eqnarray}

\begin{eqnarray}
\label{eq11}\eta=-\frac{2}{\kappa^2}\frac{V''}{V}+\frac{2}{\kappa^{2}}\frac{V'^{2}}{V^{2}}
-\frac{9}{2\kappa^{2}}\frac{\sqrt{\frac{\kappa^{2}}{3}V+\frac{1}{6}\sum_{i=1}^{3}(\dot{\xi}^{i})^2}\frac{d}{dt}\sum_{i=1}^{3}(\dot{\xi}^{i})^2}{V'^{2}}
+\frac{6\sum_{i=1}^{3}(\dot{\xi}^{i})^2}{\kappa^{2}V+\frac{1}{2}\sum_{i=1}^{3}(\dot{\xi}^{i})^2}\,.
\end{eqnarray}
To have inflationary era, these slow-roll
parameters must be very smaller than unity and when one of them
becomes unit at some point of time, the inflation ends. In the
cosmological models, duration of the inflation is given by e-folds
number $(N)$, defined as
\begin{eqnarray}
\label{eq11a}N=\int_{t_i}^{t_f} H\,dt\,,
\end{eqnarray}
with $i$ and $f$ denoting the beginning and end of this era. Now, by
using the perturbation parameters defined as~\cite{Fel11}
\begin{eqnarray}
\label{eq12}n_{s}-1=-2\epsilon-\eta\,,
\end{eqnarray}
and
\begin{eqnarray}
\label{eq13}r=16\epsilon\,,
\end{eqnarray}
We can examine how the anisotropic geometry influences the
observational viability of the model. In the next section, we
numerically analyze the model using an intermediate scale factor and
compare the results with observational data.

\section{\label{sec3}Observational Viability of the Model}
\begin{figure}
    \centering
    \includegraphics[scale=0.15, angle=0]{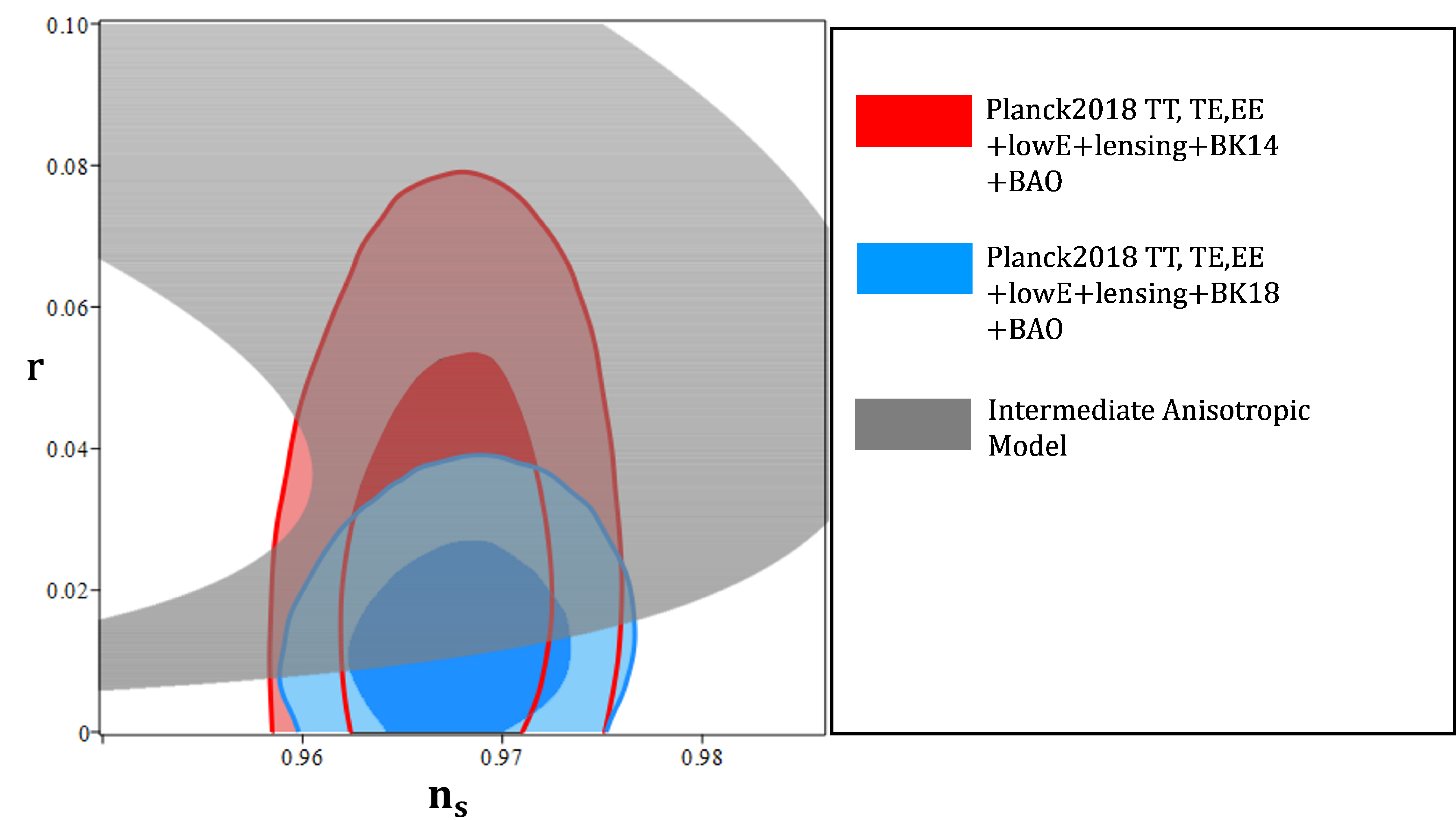}
    \caption{\label{fig1}\small {$r-n_{s}$ behavior for the anisotropic
                inflationary model with intermediate scale factor, in the background
                of Planck2018 TT, TE, EE +lowE+lensing+BK14(18) +BAO data. The
                varying parameters in this figure are $c$ and $\beta$.}}
\end{figure}

\begin{figure}
    \centering
    \includegraphics[scale=0.35, angle=0]{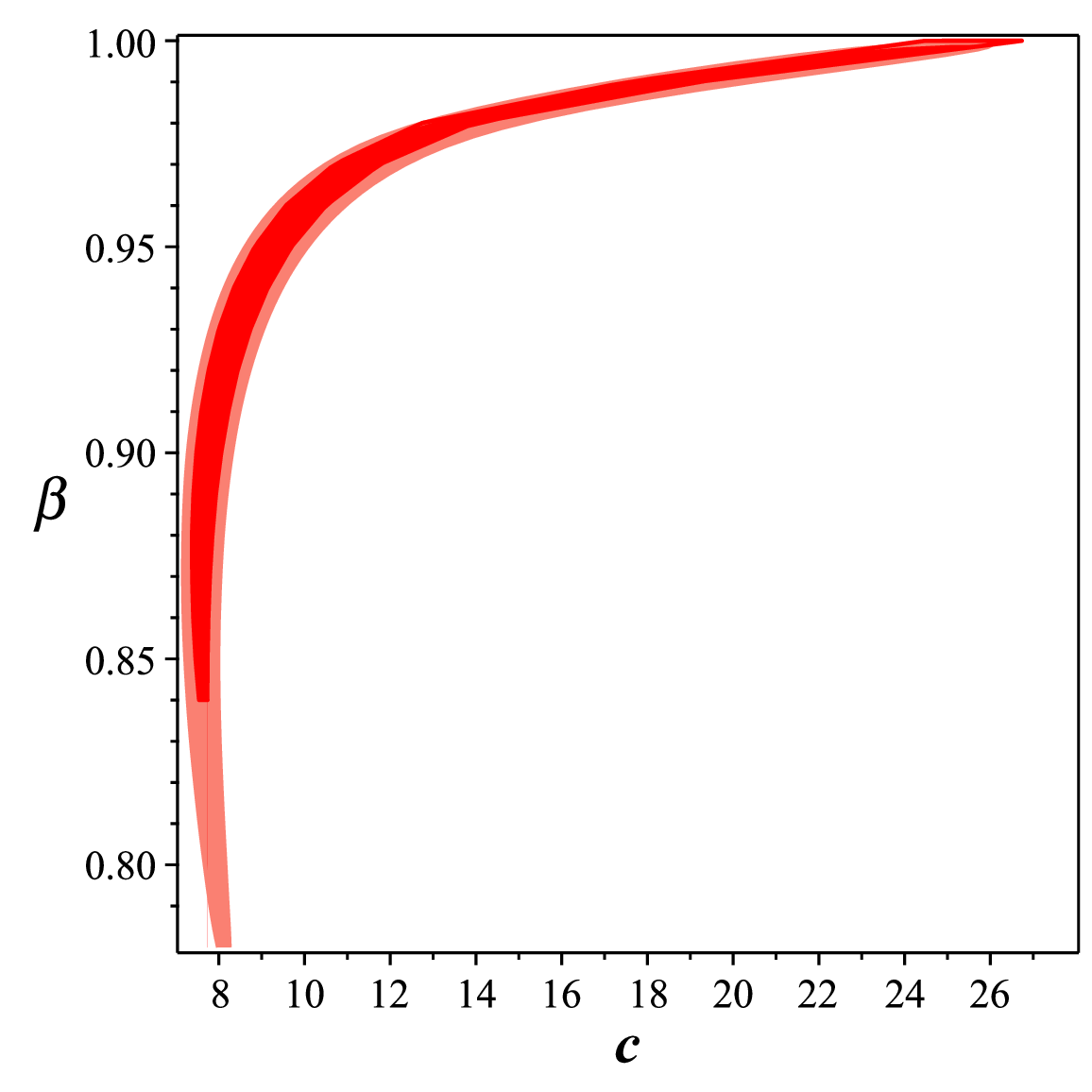}
    \includegraphics[scale=0.35, angle=0]{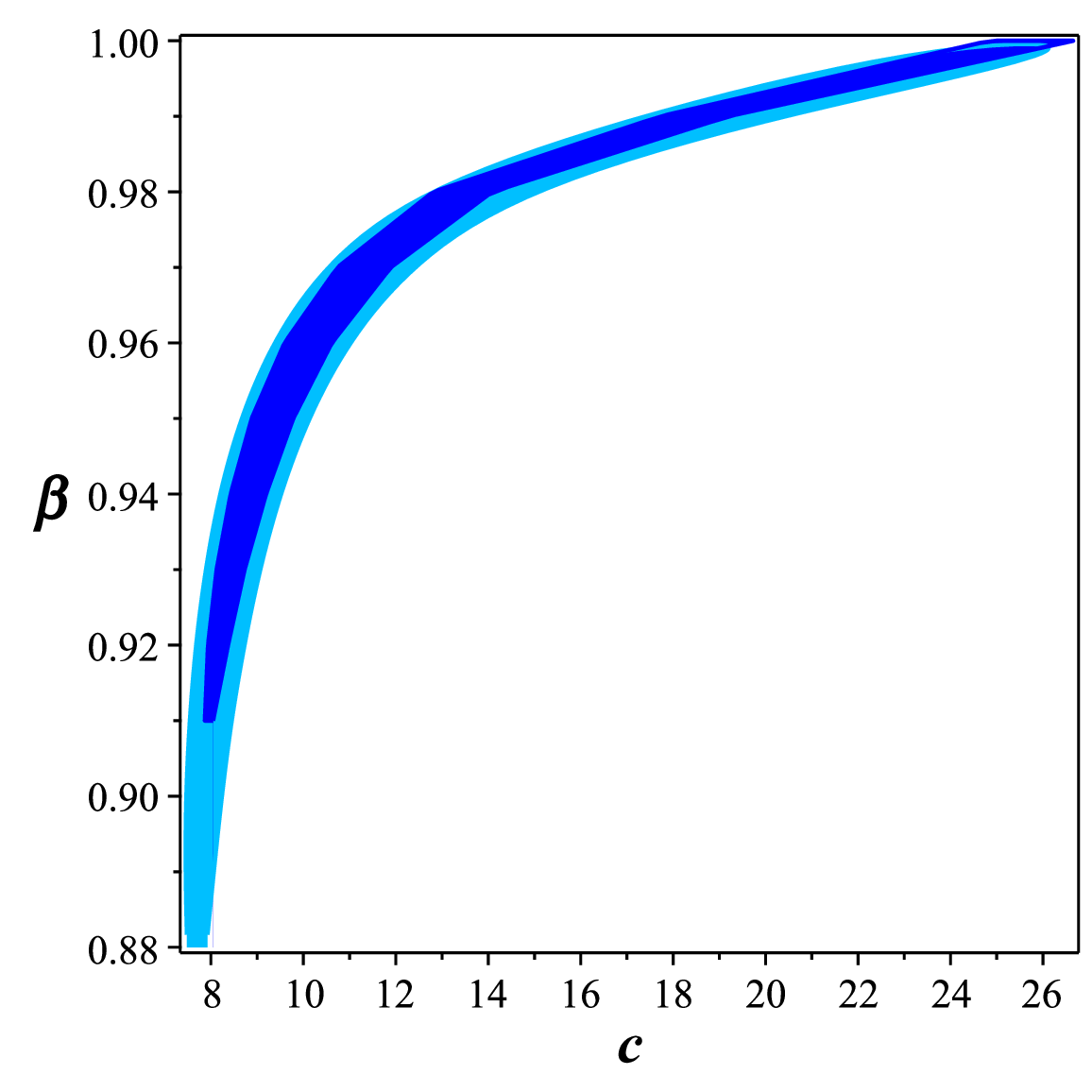}
    \caption{\label{fig2}\small {The phase space of the intermediate
            parameter $\beta$ and anisotropic parameter $c$, corresponding to
            the Planck2018 TT, TE, EE +lowE+lensing+BK14+BAO data (left panel)
            and Planck2018 TT, TE, EE +lowE+lensing+BK18+BAO data (right panel)
            at $68\%$ CL (deep colors) and $95\%$ CL (light colors). This figure
            has been plotted based on the observational constraints on the
            scalar spectral index and tensor-to-scalar ratio.}}
\end{figure}

To study the model numerically, we must find some explicit
expressions for the perturbation parameters. In this regard, we
first find some expression for $\dot{\xi}$ that appears in all our
equations. To this end, we follow Ref.~\cite{Noj22}, where it has
been shown that the parameter $\xi$ satisfies the following equation
\begin{eqnarray}
\label{eq14}\ddot{\xi}^{i}+3\,H\,\dot{\xi}^{i}=0\,.
\end{eqnarray}
Its solution gives $\dot{\xi}^{i}$ as a function of time as
\begin{eqnarray}
\label{eq15}\dot{\xi}^{i}=\frac{C^{i}}{a^{3}}\,,
\end{eqnarray}
with $C^{i}$'s to be some constants. Equation (\ref{eq15}), along
with $\Sigma_{i=1}^{3}\dot{\xi}^{i}=0$, leads to the constraint
$\Sigma_{i=1}^{3}C^{i}=0$. These equations are very important in the
numerical study of the setup. To proceed, we adopt the intermediate
scale factor defined by (see for instance~\cite{Noz18})
\begin{eqnarray}
\label{eq16}a=a_{0}\,\exp (b\,t^{\beta})\,,
\end{eqnarray}
with $b$ to be a constant and also $0<\beta<1$. By
using this equation, we can find $H$ and its derivatives in terms of
the time. Then, by using the definition of the e-folds number we
obtain the time in terms of $N$. In this way, all equations which
include $a$ or $H$ or their derivatives, become function of $N$.
This means that we can write
$H=N\left(\frac{N}{b}\right)^{-\frac{1}{\beta}}\beta$ and
$\dot{\xi}=\frac{C^{i}}{a_{0}^{3}e^{3N}}$. Then, we substitute these
definitions in equations (\ref{eq9}), (\ref{eq10}) and (\ref{eq11}),
and find the slow-roll parameters in
this setup as
\begin{eqnarray}
\label{eq17}\epsilon= \frac{\left(\frac{6 \beta^{2} b^{2}
\left(\chi^{\beta -1}\right)^{2} \left(\beta
-1\right)}{\chi}+\frac{\beta b \,\chi^{\beta -1} \left(\beta
-1\right)^{2}}{\chi^{2}}-\frac{\beta  b \,\chi^{\beta -1}
\left(\beta -1\right)}{\chi^{2}}\right)^{2}}{2 \left(-\frac{2 \beta
b \,\chi^{\beta -1} \left(\beta
-1\right)}{\chi}-\frac{c^{2}}{(a_0)^{6} \left({\mathrm e}^{b
\,\chi^{\beta}}\right)^{6}}\right) \left(3 \beta^{2} b^{2}
\left(\chi^{\beta -1}\right)^{2}+\frac{\beta  b \,\chi^{\beta -1}
\left(\beta -1\right)}{\chi}\right)^{2}}\nonumber\\+\frac{3
c^{2}}{(a_0)^{6} \left({\mathrm e}^{b \,\chi^{\beta}}\right)^{6}
\left(3 \beta^{2} b^{2} \left(\chi^{\beta -1}\right)^{2}+\frac{\beta
b \,\chi^{\beta -1} \left(\beta -1\right)}{\chi}+\frac{c^{2}}{2
(a_0)^{6} \left({\mathrm e}^{b \,\chi^{\beta}}\right)^{6}}\right)}
\,,
\end{eqnarray}
and
\begin{eqnarray}
\label{eq18}\eta = \frac{2}{3 \beta^{2} b^{2} (\chi^{\beta
-1})^{2}+\frac{Y}{\chi}} \Bigg(\frac{\kappa \bigg(-\frac{2 Y
\left(\beta -1\right)}{\chi^{2}}+\frac{2 Y}{\chi^{2}}+\frac{6 c^{2}
b \,\chi^{\beta} \beta}{(a_0)^{6} ({\mathrm e}^{b
\,\chi^{\beta}})^{6} \chi}\bigg) \left(\frac{6 \beta  b
\,\chi^{\beta -1} Y}{\chi}+\frac{Y (\beta
-1)}{\chi^{2}}-\frac{Y}{\chi^{2}}\right)}{2 \Big(-\frac{2
Y}{\chi}-\frac{c^{2}}{(a_0)^{6} ({\mathrm e}^{b
\,\chi^{\beta}})^{6}}\Big)^{\frac{5}{2}}}\nonumber\\+\frac{\frac{6
Y^{2}}{\chi^{2}}+6 \beta  b \,\chi^{\beta -1} \left(\frac{Y
\left(\beta -1\right)}{\chi^{2}}-\frac{Y}{\chi^{2}}\right)+\frac{Y
\left(\beta -1\right)^{2}}{\chi^{3}}-\frac{3 Y \left(\beta
-1\right)}{\chi^{3}}+\frac{2 Y}{\chi^{3}}}{\frac{2
Y}{\chi}+\frac{c^{2}}{(a_0)^{6} ({\mathrm e}^{b
\,\chi^{\beta}})^{6}}}\Bigg)\nonumber\\+\frac{2\Big(\frac{6 \beta b
\,\chi^{\beta -1} Y}{\chi}+\frac{Y \left(\beta
-1\right)}{\chi^{2}}-\frac{Y}{\chi^{2}}\Big)^{2}}{\big(\frac{2
Y}{\chi}+\frac{c^{2}}{(a_0)^{6} ({\mathrm e}^{b
\,\chi^{\beta}})^{6}}\big) \big(3 \beta^{2} b^{2} \left(\chi^{\beta
-1}\right)^{2}+\frac{Y}{\chi}\big)^{2}}+\frac{\, c^{2} b
\,\chi^{\beta} \beta \left(-\frac{2 Y}{\chi}-\frac{c^{2}}{(a_0)^{6}
({\mathrm e}^{b \,X^{\beta}})^{6}}\right)}{2 (a_0)^{6}
\left({\mathrm e}^{b \,\chi^{\beta}}\right)^{6} \chi \left(\frac{6
\beta b \,\chi^{\beta -1} Y}{\chi}+\frac{Y \left(\beta
-1\right)}{\chi^{2}}-\frac{Y}{\chi^{2}}\right)^{2}}\nonumber\\
\times 9 \sqrt{36 \beta^{2} b^{2} \left(\chi^{\beta
-1}\right)^{2}+\frac{12 Y}{\chi}+\frac{6 c^{2}}{(a_0)^{6} ({\mathrm
e}^{b \,\chi^{\beta}})^{6}}}\nonumber\\ +\frac{6 c^{2}}{(a_0)^{6}
\left({\mathrm e}^{b \,\chi^{\beta}}\right)^{6} \left(3 \beta^{2}
b^{2} \left(\chi^{\beta -1}\right)^{2}+\frac{\beta b \,\chi^{\beta
-1} \left(\beta -1\right)}{\chi}+\frac{c^{2}}{2 (a_0)^{6}
\left({\mathrm e}^{b \,\chi^{\beta}}\right)^{6}}\right)}\,,
\end{eqnarray}
where,
\begin{equation}
\label{eq19}\chi=\left(\frac{N}{b}\right)^{\frac{1}{\beta}}\,,\quad
Y=\beta\,b\,\chi^{\beta-1}\,(\beta-1)\,,\quad c^{2}=\sum_{i=1}^{3}
C^{i^2}\,.
\end{equation}
We substitute equations (\ref{eq17}) and (\ref{eq18}), in the
definitions of the scalar spectral index (\ref{eq12}) and
tensor-to-scalar ratio (\ref{eq13}), to perform some numerical
analysis on the model. In our analysis, we use the values of the
scalar spectral index and tensor-to-scalar ratio, released by
Planck2018 TT, TE, EE +lowE+lensing+BK14+BAO data and based on
$\Lambda$CDM+$r$+$\frac{dn_{s}}{d \ln k}$ model, as
$n_{s}=0.9658\pm0.0038$ and $r<0.072$~\cite{Pl18c}. Also, we
consider another constraint on the tensor-to-scalar ratio as
$r<0.036$, obtained from Planck2018 TT, TE, EE
+lowE+lensing+BK18+BAO data~\cite{Pa22}. In this way, in figure
\ref{fig1}, we show the evolution of $r-n_{s}$ in comparison with
Planck2018 TT, TE, EE +lowE+lensing+BK14(18)+BAO data. To plot this
figure, we have considered $7<c<30$ and $0.7<\beta<1$. In this way,
we have found constraints on $c$ using sample values of $\beta$.
These constraints are summarized in table \ref{tab1}. According to
our analysis, the model is consistent with Planck2018 TT, TE, EE
+lowE+lensing+BK14(18)+BAO data at $68\%$ CL if $0.84<\beta<1$ and
$7.70<c<27.3$ ($0.91<\beta<1$ and $8.03<c<27.2$)\footnote{The
constraints in the parentheses have been obtained from Planck2018
TT, TE, EE +lowE +lensing +BK18 +BAO data.}. The model is consistent
with the mentioned data at $95\%$ CL if $0.77<\beta<1$ and
$8.40<c<28.5$ ($0.88<\beta<1$ and $7.64<c<28.5$). The phase space of
the anisotropic parameter $c$ and intermediate parameter $\beta$,
corresponding to the $68\%$ CL and $95\%$ CL of Planck2018 TT, TE,
EE +lowE+lensing+BK14(18)+BAO constraints on $n_{s}$ and $r$, is
shown in figure \ref{fig2}. Note that, in plotting the figures, we
have adopted $N=60$ and $b=1$. According to our analysis, the
intermediate anisotropic inflation in some ranges of its model's
parameters is consistent with recent observational data. We also can
check the viability of the model from different perspectives. One
interesting perspective is the study of the reheating phase of the
universe in this model. In the following, we try to explore the
model within this context and find some more constraints on the
model's main parameters.

\begin{table*}
\tiny\tiny\caption{\small{\label{tab1} Ranges of the parameter $c$
in which the tensor-to-scalar ratio and scalar spectral index of the
intermediate anisotropic model are consistent with different data
sets.}}
\begin{center}
\tabcolsep=0.05cm\begin{tabular}{cccccc}
\\ \hline \hline \\ & Planck2018 TT,TE,EE+lowE & Planck2018 TT,TE,EE+lowE&Planck2018 TT,TE,EE+lowE&Planck2018 TT,TE,EE+lowE
\\
& +lensing+BK14+BAO &
+lensing+BK14+BAO&lensing+BK18+BAO&lensing+BK18+BAO
\\
\hline \\$\beta$& $68\%$ CL & $95\%$ CL &$68\%$ CL & $95\%$ CL
\\
\hline\hline \\  $0.8$ &  not consistent & $7.67<c<8.15$
& not consistent & not consistent\\ \\
\hline
\\$0.85$& $7.45<c<7.74$ & $7.21<c<7.97$
&not consistent& not consistent
\\ \\ \hline\\
$0.9$& $7.47<c<8.05$ & $7.27<c<8.30$
 & not consistent & $7.45<c<8.23$ \\ \\
\hline\\
$0.95$& $8.83<c<9.69$ & $8.58<c<10.02$ & $8.87<c<9.78$ &
$8.63<c<10.09$
\\ \\
\hline \hline
\end{tabular}
\end{center}
\end{table*}

\section{\label{sec4}Reheating}

\begin{figure}
    \centering
    \includegraphics[scale=0.35, angle=0]{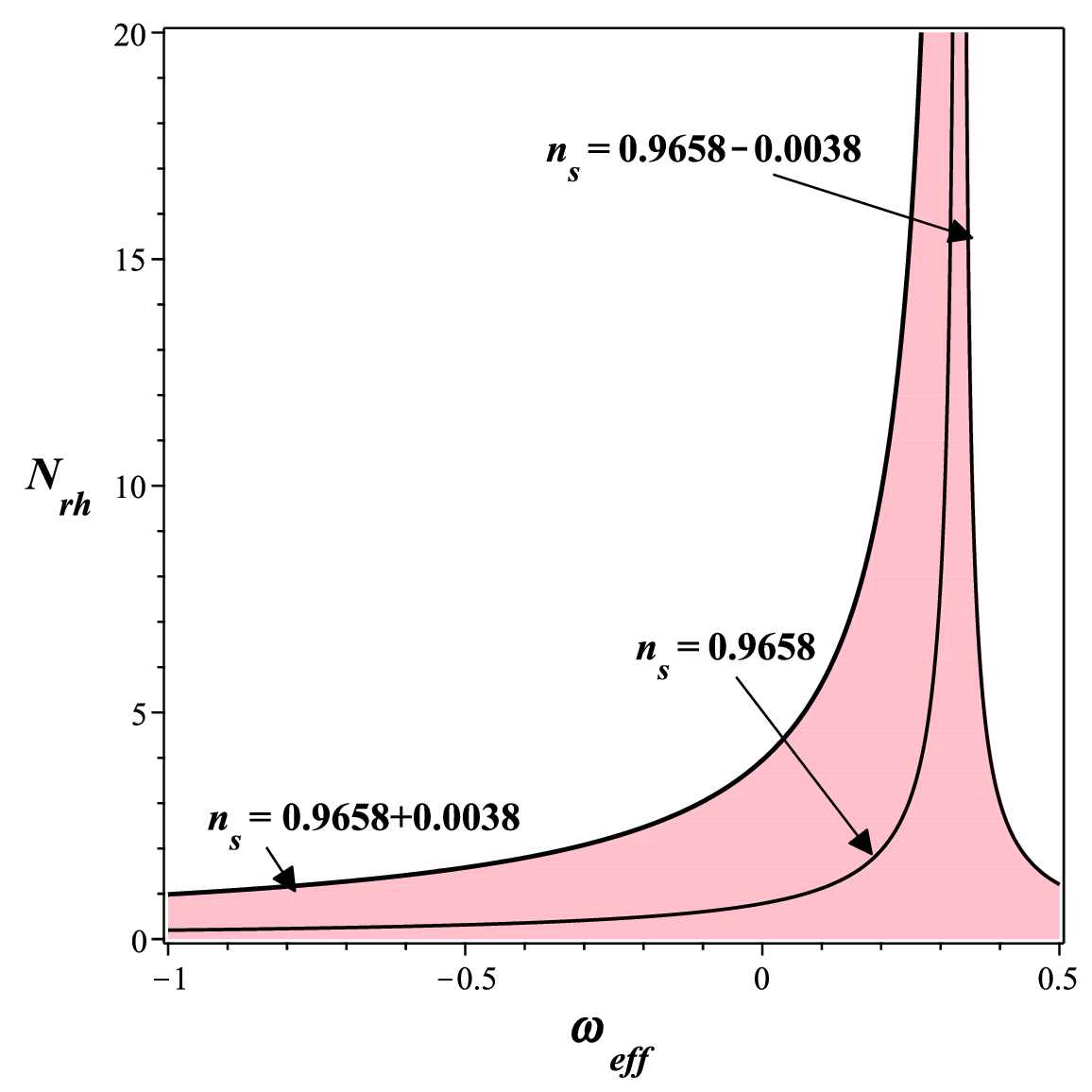}
    \includegraphics[scale=0.35, angle=0]{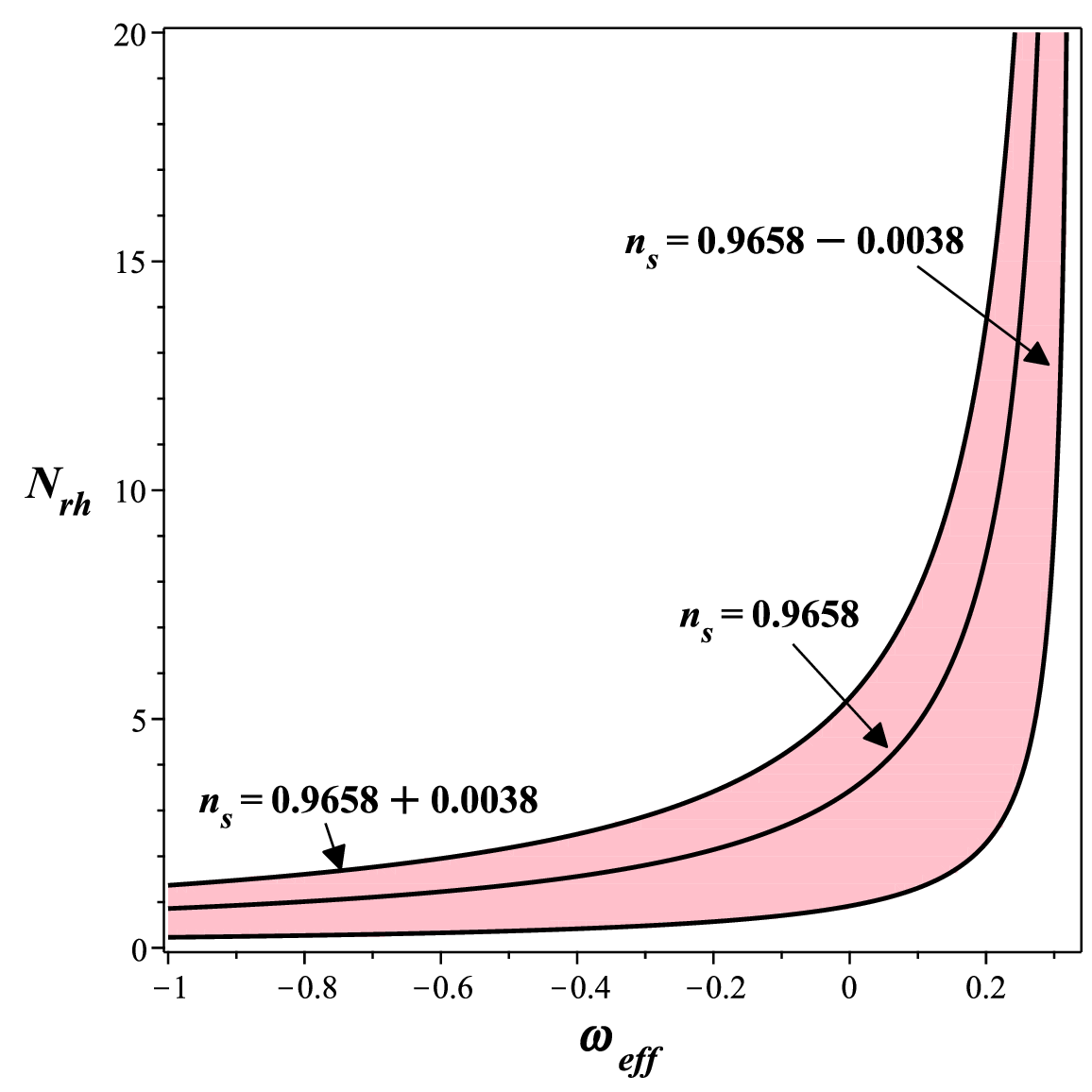}
    \caption{\label{fig3}\small {The observationally viable region of
            $N_{rh}$ and $\omega_{eff}$, based on the constraint on the scalar
            spectral index, obtained from the Planck2018 TT, TE, EE
            +lowE+lensing+BK14+BAO data. Not that, we have adopted $\beta=0.85$
            and $c=7.6$ for the left panel, and $\beta=0.95$ and $c=9.5$ for the
            right panel.}}
\end{figure}

\begin{figure}
    \centering
    \includegraphics[scale=0.35, angle=0]{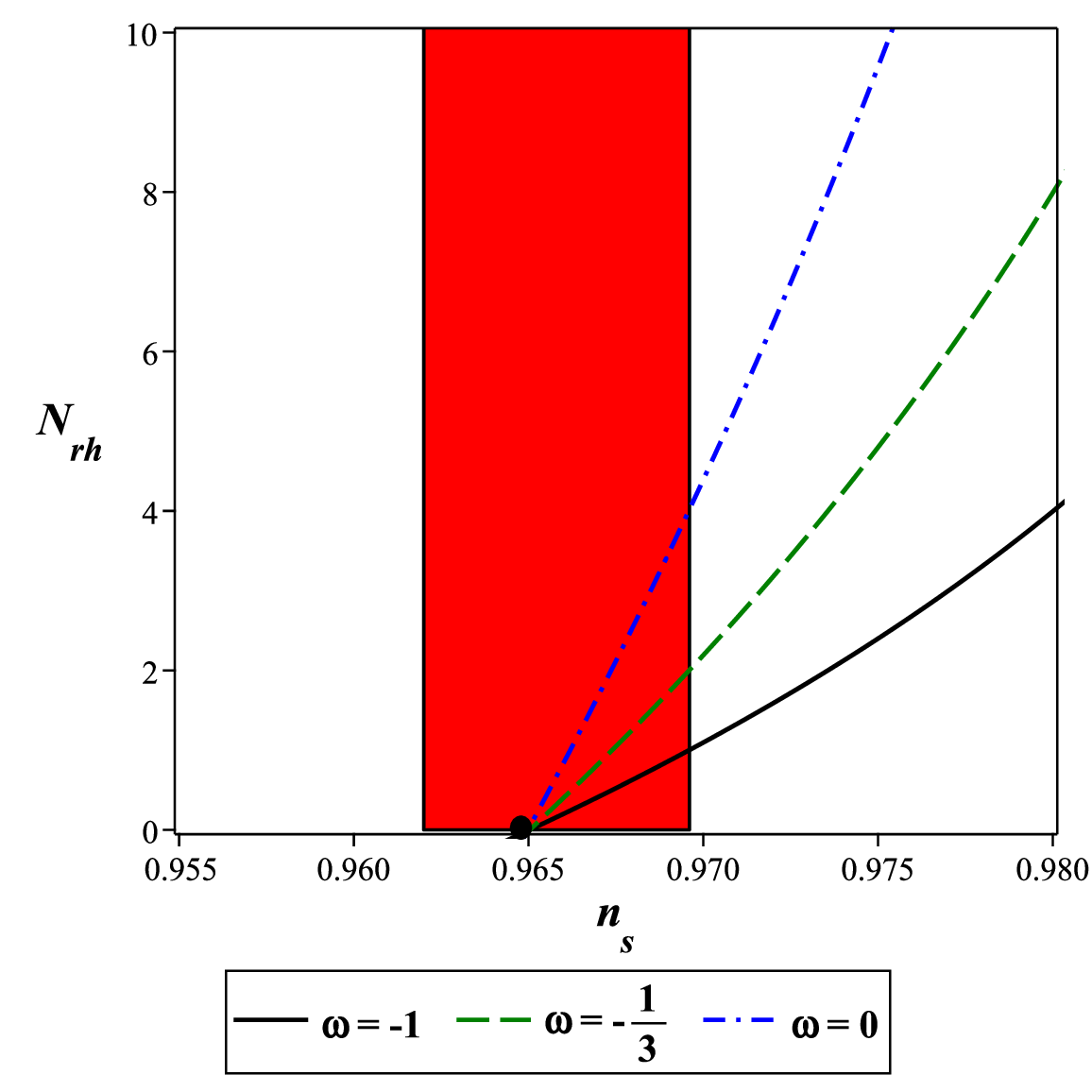}
    \includegraphics[scale=0.35, angle=0]{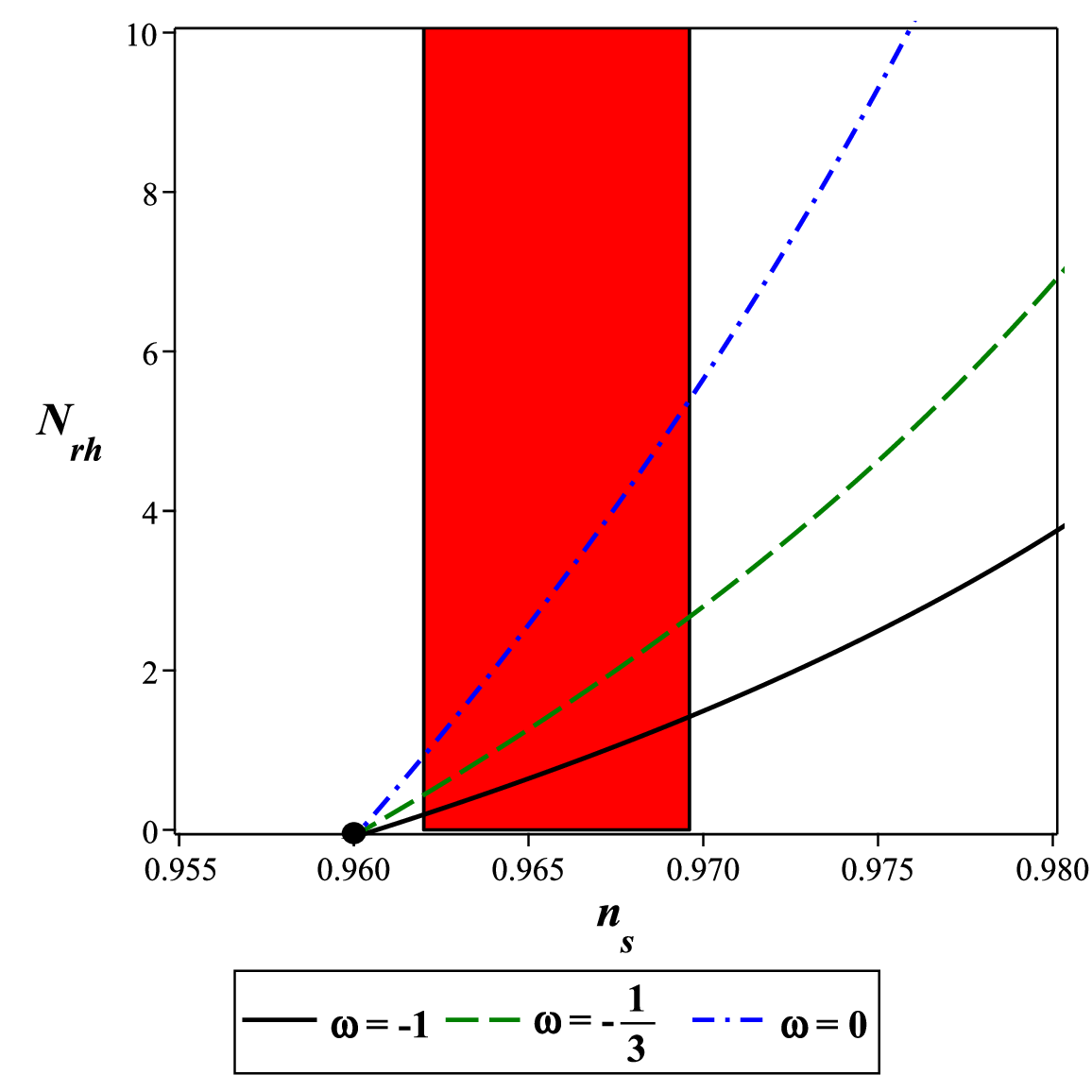}
    \caption{\label{fig4}\small {$N_{rh}-n_{s}$ behavior in the
            intermediate anisotropic model for some sample values as
            $\beta=0.85$ and $c=7.6$ (left panel), and $\beta=0.95$ and $c=9.5$
            (right panel.)}}
\end{figure}

\begin{figure}
    \centering
    \includegraphics[scale=0.35, angle=0]{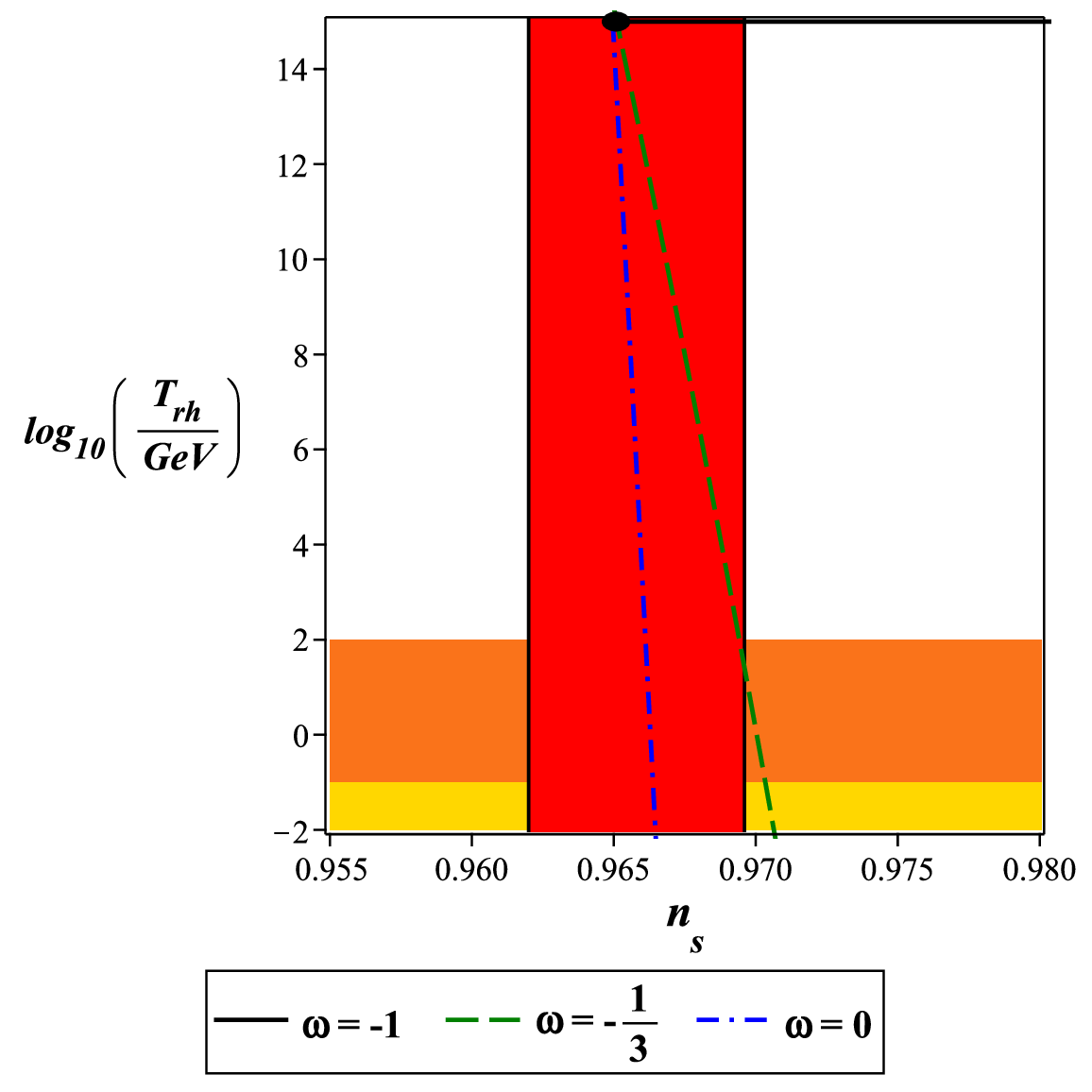}
    \includegraphics[scale=0.35, angle=0]{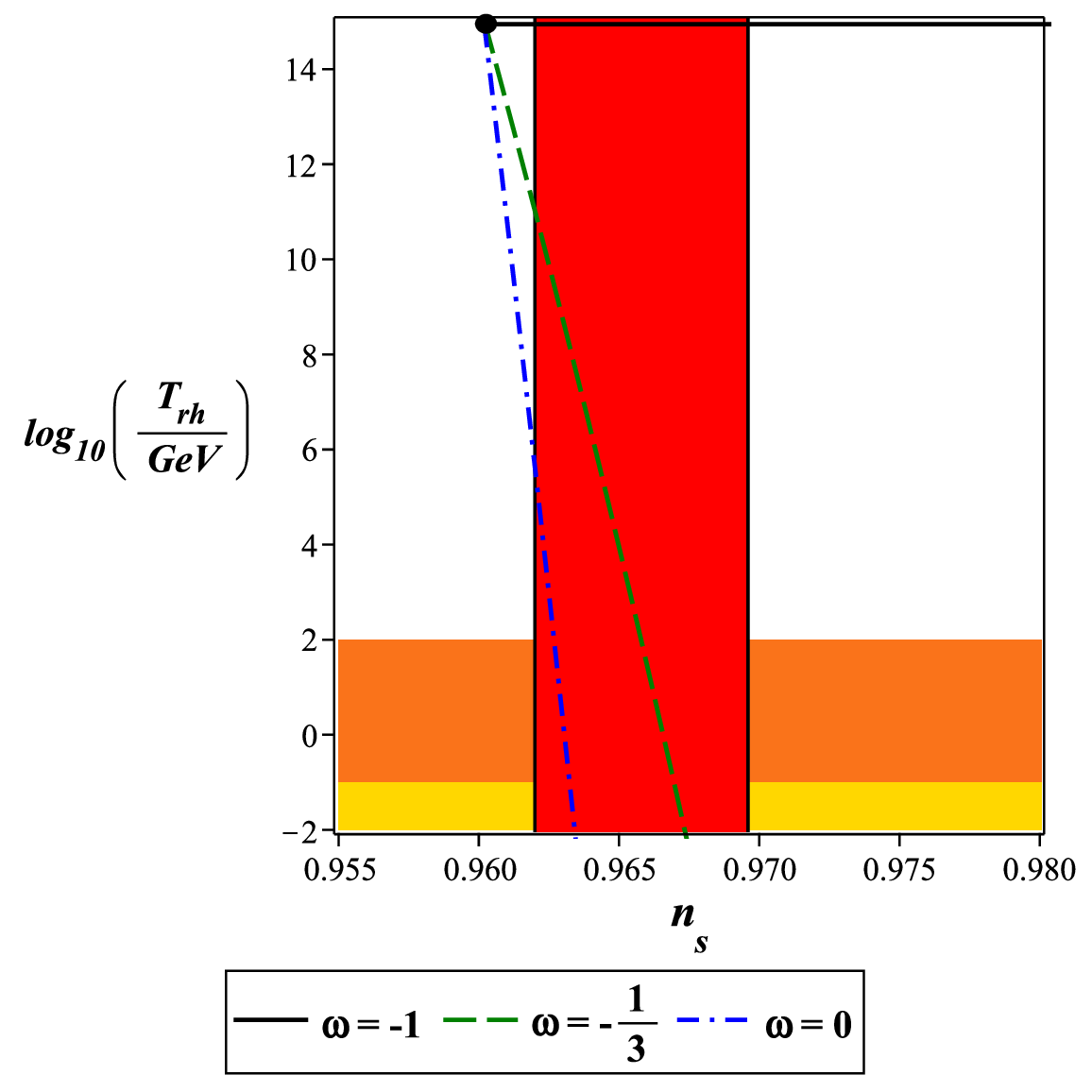}
    \caption{\label{fig5}\small {$\log_{10}\Big(\frac{T_{rh}}{GeV}\Big)-n_{s}$ behavior in the
            intermediate anisotropic model for some sample values as
            $\beta=0.85$ and $c=7.8$ (left panel), and $\beta=0.95$ and $c=9.8$
            (right panel.) The orange region corresponds to the temperatures
            below the electroweak scale, $T<100$ GeV, and the gold region stands
            for the temperatures below the big bang nucleosynthesis scale,
            $T<10$ MeV.}}
\end{figure}

After the end of the early inflationary phase of the universe, the
temperature of the universe becomes very low. Therefore the universe
couldn't be prepared to be governed by the standard model of
cosmology. It seems that, considering a reheating phase after the
end of inflation can solve this issue properly. In this regard, in ~\cite{Dai14} it hase been shown that the canonical reheating is possible in $\phi^{2}$ potential. The authors of Ref.~\cite{Un15} have constrained the equation of state parameters during the reheating. In this paper, we
use the strategy considered in the mentioned papers and also  
Res.~\cite{Co15} to examine the reheating
process in this setup (for other strategies see for
instance~\cite{Ka00}). The details of this strategy can be seen in Appendix. In this regard, we can re-write the energy density in this setup
as
\begin{eqnarray}
\label{eq27}
\rho=\frac{\left(V+\frac{1}{2\kappa^2}\sum_{i=1}^{3}(\dot{\xi}^{i})^{2}\right)}{3}\Bigg[\epsilon
-\frac{3\sum_{i=1}^{3}(\dot{\xi}^{i})^{2}}{\kappa^2
V+\frac{1}{2}\sum_{i=1}^{3}(\dot{\xi}^{i})^{2}}\Bigg] +V\,.
\end{eqnarray}
To obtain this equation, we have used the equation of
motion (\ref{eq8}) and obtained $\dot{\phi}$ in terms of the
potential and its derivative. Then, we have used equation
(\ref{eq11}) and found the $V'$ in terms of the slow-roll parameter
$\epsilon$. After that we substitute these in the energy density of
the scalar field $\Big(\rho=\frac{1}{2}\dot{\phi}^{2}+V(\phi)\Big)$
and reach the above equation for energy density. When we take
$\epsilon=1$, equation (\ref{eq27}) gives us the
energy density at the end of inflation as
\begin{eqnarray}
\label{eq28}
\rho_{e}=\frac{\left(V_{e}+\frac{1}{2\kappa^2}\sum_{i=1}^{3}(\dot{\xi}^{i})^{2}\right)}{3}\Bigg[1
-\frac{3\sum_{i=1}^{3}(\dot{\xi}^{i})^{2}}{\kappa^2
V_{e}+\frac{1}{2}\sum_{i=1}^{3}(\dot{\xi}^{i})^{2}}\Bigg] +V_{e}\,.
\end{eqnarray}
leading to
\begin{eqnarray}
\label{eq29}
\rho_{rh}=\Bigg[\frac{\left(V_{e}+\frac{1}{2\kappa^2}\sum_{i=1}^{3}(\dot{\xi}^{i})^{2}\right)}{3}\Bigg(1
-\frac{3\sum_{i=1}^{3}(\dot{\xi}^{i})^{2}}{\kappa^2
V_{e}+\frac{1}{2}\sum_{i=1}^{3}(\dot{\xi}^{i})^{2}}\Bigg)
+V_{e}\Bigg]\,\exp\Big[-3N_{rh}(1+\omega_{eff})\Big].
\end{eqnarray}
Now, with equation (\ref{eq29}) we can write
\begin{eqnarray}
\label{eq30}
\ln\left(\frac{a_{0}}{a_{rh}}\right)=-\frac{1}{3}\ln\left(\frac{43}{11g_{rh}}\right)
-\frac{1}{4}\ln\left(\frac{\pi^{2}g_{rh}}{30}\right)-\ln
T_{0}-\frac{3}{4}N_{rh}(1+\omega_{eff}) \nonumber\\
+\frac{1}{4}\ln\Bigg(\frac{\left(V_{e}+\frac{1}{2\kappa^2}\sum_{i=1}^{3}(\dot{\xi}^{i})^{2}\right)}{3}\Bigg[1
-\frac{3\sum_{i=1}^{3}(\dot{\xi}^{i})^{2}}{\kappa^2
V_{e}+\frac{1}{2}\sum_{i=1}^{3}(\dot{\xi}^{i})^{2}}\Bigg]
+V_{e}\Bigg)\,.
\end{eqnarray}
With this equation, we can find the
number of e-folds during the reheating in terms of the anisotropic
and other model's parameters as
\begin{eqnarray}
\label{eq31}
N_{rh}=\frac{4}{1-3\omega_{eff}}\Bigg[-N-\ln\left(\frac{k_{hc}}{a_{0}T_{0}}\Big(\frac{30}{\pi^{2}g_{rh}}\Big)^{\frac{1}{4}}\Big(\frac{11g_{rh}}{43}\Big)^{\frac{1}{3}}\right)
+\frac{1}{2}\ln(H^{2})\nonumber\\
-\frac{1}{4}\ln\Bigg(\frac{\left(V_{e}+\frac{1}{2\kappa^2}\sum_{i=1}^{3}(\dot{\xi}^{i})^{2}\right)}{3}\Bigg(1
-\frac{3\sum_{i=1}^{3}(\dot{\xi}^{i})^{2}}{\kappa^2
V_{e}+\frac{1}{2}\sum_{i=1}^{3}(\dot{\xi}^{i})^{2}}\Bigg)
+V_{e}\Bigg)\Bigg],
\end{eqnarray}
leading to the following expression for the reheating temperature
\begin{equation}
\label{eq32}
T_{rh}=\bigg(\frac{30}{\pi^{2}g_{rh}}\bigg)^{\frac{1}{4}}\,
\Bigg[\frac{\left(V_{e}+\frac{1}{2\kappa^2}\sum_{i=1}^{3}(\dot{\xi}^{i})^{2}\right)}{3}\Bigg(1
-\frac{3\sum_{i=1}^{3}(\dot{\xi}^{i})^{2}}{\kappa^2
V_{e}+\frac{1}{2}\sum_{i=1}^{3}(\dot{\xi}^{i})^{2}}\Bigg)
+V_{e}\Bigg]^{\frac{1}{4}}\,\exp\bigg[-\frac{3}{4}N_{rh}(1+\omega_{eff})\bigg]\,.
\end{equation}
Now, we have enough tools to explore the observational viability of
the model in the context of the reheating process and by using the
important parameters $N_{rh}$ and $T_{rh}$. There are some points
that should be taken care to be ready for numerical analysis. In
equations (\ref{eq31}) and (\ref{eq32}), we see the parameter
$V_{e}$. We can use the fact that at the end of inflation, we have
$\epsilon=1$. From this condition and equation (\ref{eq17}), we find
the parameter $c$ in terms of other parameters (this gives us a very
long-expression and we avoid presenting it here). Then, we
substitute it in equation (\ref{eq9}) to find the value of the
potential at the end of the inflation. Note that, to find the full
expression in terms of the model's constant parameters, we should
use equations (\ref{eq15}) and (\ref{eq16}) too. Since we want to
study the reheating parameter in comparison with observational data,
we should have them in terms of some parameters that are constrained
with data. We can write $H^2$ in the equation (\ref{eq31}) in terms
of the scalar spectral index. To this end, we consider the
definition of the first slow-roll parameter as
$\epsilon=-\frac{\dot{H}}{H^{2}}$. Then, we find $H^{2}=-\epsilon
\dot{H}$, where $\dot{H}=N^{\frac{\beta -2}{\beta}} \beta
\,b^{\frac{2}{\beta}} \left(\beta -1\right)$ in the intermediate
model with $a=a_{0}\exp(bt^{\beta})$. On the other hand, we have
$n_{s}=1-2\epsilon-\eta$, leading to
$\epsilon=\frac{1}{2}(1-n_{s}-\eta)$. By substituting this recent
equation in $H^{2}=-\epsilon \dot{H}$, we find
\begin{equation}
H^{2}=-\frac{1}{2}\Big[1-n_{s}-\eta\Big]\left[N^{\frac{\beta -2}{\beta}} \beta  \,b^{\frac{2}{\beta}} \left(\beta -1\right)\right]\,,
\end{equation}
where $\eta$ is given by equation (\ref{eq18}). In this way, we
obtain equations (\ref{eq31}) and therefore (\ref{eq32}) in terms of
the scalar spectral index and perform some numerical analysis. We
use the observational value of the scalar spectral index,
$n_{s}=0.9658\pm 0.0038$, obtained from Planck2018 TT, TE, EE
+lowE+lensing+BK14+BAO data. We also set
$g_{rh}=106.75$~\cite{Cre14} and $k_{hc}=0.002\,Mpc^{-1}$ as pivot
scale~\cite{Pl18c}. By using these values, we obtain the
observationally viable ranges of $N_{rh}$ and $\omega_{eff}$ and
results are shown in figure \ref{fig3}. In this figure, we have
chosen two values for the intermediate and anisotropic parameters as
$\beta=0.85,\, c=7.6$ and $\beta=0.95,\, c=9.5$. From the figure, it
seems that for $\beta=0.85,\, c=8$ it is possible to have
instantaneous reheating. However, for $\beta=0.95,\, c=9$, the
universe reheats after a few numbers of e-folds. Since at the end of
inflation era, the value of the effective equation of state is
$-\frac{1}{3}$ and at the beginning of the radiation dominated era
it is $\frac{1}{3}$, it is important for the model to cover this
transition. According to the figure \ref{fig3}, it seems that by
increasing the e-folds number during the reheating, the effective
equation of state parameter increases from $-1$ to $\frac{1}{3}$
where the universe becomes radiation dominated. We have also plotted
$N_{rh}-n_{s}$ and $T_{rh}-n_{s}$ behaviors for these sample values
and for $\omega=-1$, $\omega=-\frac{1}{3}$ and $\omega=0$ that have
been shown in figures \ref{fig4} and \ref{fig5}. By considering the
observationally viable ranges of the anisotropic parameter, based on
the observational constraints on $n_{s}$ and $r$, we have obtained
some constraints on the amount of the e-folds number and
temperature, needed for the universe to be reheated. The results are
shown in tables \ref{tab2} and \ref{tab3}. The results are in both
$68\%$ CL and $95\%$ CL. These data also show that in some regions
of the model's parameters, it is possible to have instantaneous
reheating phase. However, for some other regions, some e-folds
needed in the reheating process. According to our analysis and in
the ranges of the considered anisotropic and intermediate
parameters, it seems that for $\omega=-1$, depending on the values
of $\beta$ and $c$, the constraint on the e-folds number and
temperature during reheating are respectively as $0\leq N_{rh} \leq
1.80$ and $14.97\leq \log_{10}\left(\frac{T_{rh}}{GeV}\right)\leq
15$ at $68\%$ CL, and $0\leq N_{rh} \leq 2.15$ and $14.96\leq
\log_{10}\left(\frac{T_{rh}}{GeV}\right)\leq 15$ at $95\%$ CL. These
constraints for $\omega=-\frac{1}{3}$ are as $0\leq N_{rh} \leq
2.70$ and $-1.92\leq \log_{10}\left(\frac{T_{rh}}{GeV}\right)\leq
15$ at $68\%$ CL, and $0\leq N_{rh} \leq 3.50$ and $-1.71\leq
\log_{10}\left(\frac{T_{rh}}{GeV}\right)\leq 15$ at $95\%$ CL. Also,
for $\omega=0$ we have $0\leq N_{rh} \leq 5.85$ and $-1.89\leq
\log_{10}\left(\frac{T_{rh}}{GeV}\right)\leq 15$ at $68\%$ CL, and
$0\leq N_{rh} \leq 6.13$ and $-1.90\leq
\log_{10}\left(\frac{T_{rh}}{GeV}\right)\leq 15$ at $95\%$ CL.

\begin{table*}
\tiny \tiny \caption{\small{\label{tab2} Constraints on $N_{rh}$ in
the anisotropic inflationary model with intermediate scale factor,
based on the constraint on the scalar spectral index and
tensor-to-scalar ratio, obtained from Planck2018 TT, TE,
EE+lowE+lensing+BK14+BAO joint data at $68\%$ CL and $95\%$ CL.}}
\begin{center}
\begin{tabular}{ccccccc}
\\ \hline \hline \\ &$\beta$&$c$& $\omega=-1$& $\omega=-\frac{1}{3}$
&$\omega=0$
\\
\hline\\
&$0.85$&$7.45<c<7.74$&$0\leq N_{rh}\leq1.33$&$0\leq N_{rh}\leq2.27$&$0\leq N_{rh}\leq4.27$\\ \\
$68\%$ CL&$0.90$&$7.47<c<8.05$&$0.04\leq N_{rh}\leq1.57$&$0.10\leq N_{rh}\leq2.43$&$0.72\leq N_{rh}\leq4.41$\\ \\
&$0.95$&$8.83<c<9.69$&$0.10\leq N_{rh}\leq1.80$&$0.17\leq
N_{rh}\leq2.70$&$0.78\leq N_{rh}\leq5.85$\\ \\ \hline\\
&$0.85$&$7.21<c<7.97$&$0\leq N_{rh}\leq1.50$&$0\leq N_{rh}\leq2.35$&$0\leq N_{rh}\leq4.39$\\ \\
$95\%$ CL&$0.90$&$7.27<c<8.30$&$0\leq N_{rh}\leq1.70$&$0.05\leq N_{rh}\leq2.53$&$0.64\leq N_{rh}\leq4.63$\\ \\
&$0.95$&$8.58<c<10.02$&$0.08\leq N_{rh}\leq2.15$&$0.11\leq
N_{rh}\leq3.50$&$0.74\leq N_{rh}\leq6.13$\\ \\ \hline \hline
\end{tabular}
\end{center}
\end{table*}

\begin{table}
\tiny \tiny \caption{\small{\label{tab3} Constraints on the $T_{rh}$
in the anisotropic inflationary model with intermediate scale
factor, based on the constraint on the scalar spectral index and
tensor-t-scalar ratio, obtained from Planck2018 TT, TE,
EE+lowE+lensing+BK14+BAO joint data at $68\%$ CL and $95\%$ CL.}}
\begin{center}
\begin{tabular}{ccccccc}
\\ \hline \hline \\ &$\beta$&$c$& $\omega=-1$& $\omega=-\frac{1}{3}$
&$\omega=0$
\\
\hline\\
&$0.85$&$7.45<c<7.74$&$14.98\leq
\log_{10}\left(\frac{T_{rh}}{GeV}\right)\leq15.00$&$1.14\leq
\log_{10}\left(\frac{T_{rh}}{GeV}\right)\leq15$&$-1.90\leq
\log_{10}\left(\frac{T_{rh}}{GeV}\right)\leq15$\\ \\
$68\%$ CL&$0.90$&$7.47<c<8.05$&$14.98\leq
\log_{10}\left(\frac{T_{rh}}{GeV}\right)\leq15.00$&$0.06\leq
\log_{10}\left(\frac{T_{rh}}{GeV}\right)\leq14.37$&$-1.89\leq
\log_{10}\left(\frac{T_{rh}}{GeV}\right)\leq8.87$\\ \\
&$0.95$&$8.83<c<9.69$&$14.97\leq
\log_{10}\left(\frac{T_{rh}}{GeV}\right)\leq14.99$&$-1.92\leq
\log_{10}\left(\frac{T_{rh}}{GeV}\right)\leq9.89$&$-1.89\leq
\log_{10}\left(\frac{T_{rh}}{GeV}\right)\leq3.14$\\ \\
\hline\\
&$0.85$&$7.21<c<7.97$&$14.97\leq
\log_{10}\left(\frac{T_{rh}}{GeV}\right)\leq15.00$&$1.02\leq
\log_{10}\left(\frac{T_{rh}}{GeV}\right)\leq15$&$-1.91\leq
\log_{10}\left(\frac{T_{rh}}{GeV}\right)\leq15$\\ \\
$95\%$ CL&$0.90$&$7.27<c<8.30$&$14.96\leq
\log_{10}\left(\frac{T_{rh}}{GeV}\right)\leq15.00$&$-1.80\leq
\log_{10}\left(\frac{T_{rh}}{GeV}\right)\leq14.65$&$-1.90\leq
\log_{10}\left(\frac{T_{rh}}{GeV}\right)\leq10.10$\\ \\
&$0.95$&$8.58<c<10.02$&$14.96\leq
\log_{10}\left(\frac{T_{rh}}{GeV}\right)\leq15.00$&$-1.71\leq
\log_{10}\left(\frac{T_{rh}}{GeV}\right)\leq10.74$&$-1.90\leq
\log_{10}\left(\frac{T_{rh}}{GeV}\right)\leq5.32$\\ \\
 \hline \hline
\end{tabular}
\end{center}
\end{table}

\section{\label{sec5}Conclusion}
In this paper, we revisit intermediate inflation within an
anisotropic geometry. By incorporating the Friedmann equations
expressed in terms of anisotropic parameters, we derive the
slow-roll parameters for this framework. Since the anisotropic
property appears directly in the Friedmann equations, its effects
appear in the slow-roll parameters too. This, consequently, affects
the perturbations parameters scalar spectral index and
tensor-to-scalar ratio. Then, in order to check the viability of the
model, we have adopted the intermediate scale factor as $a=a_{0}\exp
(bt^{\beta})$. We have also considered the anisotropic parameter
$\dot{\xi}^{i}=\frac{C^{i}}{a^{3}}$. By these choices, we have
obtained the slow-roll parameters and therefore, the perturbations
parameters in terms of the intermediate $(\beta)$ and anisotropic
$(c)$ parameters. In this way, we have tried to perform some
numerical analysis and find some constraints on $\beta$ and $c$. To
this end, we have considered both Planck2018 TT, TE,
EE+lowE+lensing+BK14+BAO and Planck2018 TT, TE,
EE+lowE+lensing+BK18+BAO data. We have studied $r-n_{s}$ behavior in
comparison to these data and found some constraints on the model's
parameters. We have also plotted the $\beta-c$ phase space at both
$68\%$ CL and $95\%$ CL. The constraints that we have obtained from
our analysis are $0.84<\beta<1$ and $7.34<c<27.7$ ($0.91<\beta<1$
and $8.00<c<27.4$) based on Planck2018 TT, TE, EE
+lowE+lensing+BK14(18)+BAO data at $68\%$ CL. Also, we have obtained
$0.77<\beta<1$ and $7.17<c<28.9$ ($0.88<\beta<1$ and
$7.40<c<28.8$), based on Planck2018 TT, TE, EE
+lowE+lensing+BK14(18)+BAO data at $95\%$ CL. After that, we studied
the reheating phase after inflation in the intermediate anisotropic
model. We have obtained some important parameters such as the
e-folds number and temperature parameters in terms of $\beta$ and
$c$. Then, we have used the observational constraints on $\beta$ and
$c$, to explore the viable ranges of the reheating parameters. In
this regard, we have plotted the regions of $N_{rh}$ and $T_{rh}$
versus $\omega_{eff}$ that are observationally viable. We have shown
that when we use the observationally viable ranges of the model's
parameters, it is possible to have a viable reheating phase.

In summary, in the absence of the anisotropic effect, the
intermediate inflation with a simple single scalar field is not
consistent with new observational data. This point has been shown in
figure \ref{fig2}, where the observationally viable ranges of $c$
start from $c=7$. Note that, this result holds even with $\beta=1$,
corresponding to pure exponential scale factor. In fact, the
presence of the anisotropic leads to a shift in both the scalar
spectral index and tensor-to-scalar ratio and makes them
observationally viable. Therefore, it seems that the anisotropic
geometry has an important role in the viability of the model. Also,
the viable ranges of $c$ and $\beta$ is important in studying the
reheating phase. As figure \ref{fig3} shows, when the value of
anisotropic parameter is larger, the instantaneous reheating is not
possible (the right panel of figure \ref{fig3}). Also, for smaller
values of $c$, the reheating process can happen instantaneously. In
this regard, the value of $c$ in the model seems important in the
sense that it determines the type of reheating phase. With the
obtained ranges of the anisotropic parameter, the model supports the
big-bang nucleosynthesis. One important result in
studying the reheating phase in oue model is that by increasing the
e-folds number during the reheating, the effective equation of state
parameter increases from $-1$ to $\frac{1}{3}$ where the universe
becomes radiation dominated.
\\

\textbf{ACKNOWLEDGMENTS}

We thank the respected referees for the very insightful comments that have
improved the quality of the paper considerably.
\\

\textbf{Appendix}

Following Refs.~\cite{Ue16}, we consider the
following relation for the e-folds number
\begin{equation}
    \label{eq20} N_{hc}=\ln \Bigg(\frac{a_{end}}{a_{hc}}\Bigg)\,.
\end{equation}
Note that this parameter is defined in the interval between the time
when physical scales crosses the Hubble horizon (shown by $hc$) and
the time when inflation ends (shown by $end$). In the next step, it
is important to have the e-folds number during the
reheating phase $(N_{rh})$ in terms of the
effective equation of state ($\omega_{rh}$) and energy density
during the reheating. To do this, we can use $\rho\sim
a^{-3(1+\omega_{eff})}$. Therefore, we obtain~\cite{Ue16}
\begin{eqnarray}
    \label{eq21}
    N_{rh}=\ln\Bigg[\frac{a_{rh}}{a_{end}}\Bigg]=-\frac{1}{3(1+\omega_{eff})}\ln\Bigg[\frac{\rho_{rh}}{\rho_{end}}\Bigg]\,.
\end{eqnarray}
We also have the following relation between the parameters at the
time when the physical scales cross the Hubble horizon
\begin{eqnarray}
    \label{eq22} 0=\ln\Bigg[\frac{k_{hc}H_{hc}^{-1}}{a_{hc}}\Bigg]\,,
\end{eqnarray}
leading to
\begin{eqnarray}
    \label{eq23}
    N_{hc}+N_{rh}+\ln\Bigg[\frac{k_{hc}H_{hc}^{-1}}{a_{0}}\Bigg]+\ln\Bigg[\frac{a_{0}}{a_{rh}}\Bigg]=0\,.
\end{eqnarray}
Now, we consider the following
equation~\cite{Ue16}
\begin{equation}
    \label{eq24} \rho_{rh}=\frac{\pi^{2}g_{rh}}{30}T_{rh}^{4}\,,
\end{equation}
which gives us the relation between the energy density and
temperature during the reheating. In this equation, the parameter
$g_{rh}$ stands for the effective number of the relativistic species
at the reheating phase. We have also the relation between the scale
factor and temperature as~\cite{Co15,Ue16}
\begin{equation}
    \label{eq25}
    a_{0}=\left(\frac{11g_{rh}T_{rh}^{3}}{43T_{0}^{3}}\right)^{\frac{1}{3}}a_{rh}\,.
\end{equation}
Equations (\ref{eq24}) and (\ref{eq25}) give~\cite{Ue16}
\begin{eqnarray}
    \label{eq26}
    \frac{a_{0}}{a_{rh}}=\left(\frac{11g_{rh}}{43T_{0}^{3}}\right)^{\frac{1}{3}}\left(\frac{30\rho_{rh}}{\pi^{2}g_{rh}}\right)^{\frac{1}{4}}\,.
\end{eqnarray}
This is a useful equation in studying the reheating phase.

\end{document}